\documentclass{appolb}
\usepackage{epsfig}
\usepackage{color}
\usepackage{cite}
\begin{document}
\pagestyle{plain}
\newcount\eLiNe\eLiNe=\inputlineno\advance\eLiNe by -1
\title{Dynamic structural and topological phase transitions on the Warsaw Stock Exchange: A phenomenological approach}
\author{A. Sienkiewicz,
%\footnote{email: asien@okwf.edu.pl},
%\address{Institute of Experimental Physics, Faculty of Physics \\ University of Warsaw, 
%Ho\.za 69, PL-00681 Warsaw, Poland} 
T. Gubiec,
%\footnote{email: Tomasz.Gubiec@fuw.edu.pl},
%\address{Institute of Experimental Physics, Faculty of Physics \\ University of Warsaw, 
%Ho\.za 69, PL-00681 Warsaw, Poland}
R. Kutner\footnote{email: Ryszard.Kutner@fuw.edu.pl}
\address{Institute of Experimental Physics \\ Faculty of Physics, University of Warsaw \\ 
Ho\.za 69, PL-00681 Warsaw, Poland} \\
\& \\
Z. R. Struzik
%\footnote{email: zbigniew.struzik@p.u-tokyo.ac.jp}
\address{The University of Tokyo, Bunkyo-ku, Tokyo 113-0033, Japan}
}
\maketitle

\begin{abstract}
We study the crash dynamics of the Warsaw Stock Exchange (WSE) by using the Minimal Spanning Tree (MST) networks. 
We find the transition of the complex network during its evolution from a (hierarchical) power law MST network, 
representing the stable state of WSE before the recent worldwide financial crash, 
to a superstar-like (or superhub) MST network of the market decorated by a hierarchy of trees 
(being, perhaps, an unstable, intermediate market state). Subsequently, we observed 
a transition from this complex tree to the topology of the (hierarchical) power law MST network decorated by several 
star-like trees or hubs. 
This structure and topology represent, perhaps, the WSE after the worldwide financial crash, and could be 
considered to be an aftershock. 
Our results can serve as an empirical foundation for a future theory of dynamic structural and topological phase 
transitions on financial markets.  
\end{abstract}
%\maketitle

\section{Introduction}\label{section:hypothesis}

It is only in the past two decades that physicists have intensively studied the structural and/or topological 
properties of complex networks \cite{DGM} (and refs. therein). They have discovered that in most real graphs, 
small and finite loops are rare and insignificant. Hence, it was possible to assume their architectures to be 
locally dominated by trees. These properties have been extensively exploited. For instance, it is surprising how 
well this assumption works in the case of numerous loopy and clustered 
networks\footnote{Nevertheless, the stability problem of the networks versus their structure and topology should 
be studied.}.
Therefore, we decided on the Minimal Spanning Tree (MST) technique as a particularly useful, canonical tool in 
graph theory \cite{BeBo}, being a correlation based connected network without any loop 
\cite{RNM,BCLMVM,MS,BLM,VBT,KKK,TMAM,TCLMM} (and refs. therein). In the graph, the vertices (nodes) are the 
companies and the distances between them are obtained from the corresponding correlation coefficients. The 
required transformation of the correlation coefficients into distances was made according to the simple recipe 
\cite{RNM,MS}. 

We consider the dynamics of an empirical complex network of companies, which were listed on 
the Warsaw Stock Exchange (WSE) for the entire duration of each period of time in question. In general, both 
the number of companies (vertices) and distances between them can vary in time. That is, in a given period 
of time these quantities are fixed but in other periods can be varied. Obviously, during the network evolution 
some of its edges may disappear, while others may emerge. Hence, neither the number of companies nor edges are 
conserved quantities. As a result, their characteristics, such as for instance, their mean length and mean 
occupation layer \cite{RNM,BLM,BR,TSC,OCKK1,OCKK2,OCKKK}, are continuously varying over time as discussed below.

We applied the MST technique to find the transition of a complex network during its evolution from a hierarchical 
(power law) tree representing the stock market structure before the recent worldwide financial crash \cite{DiSo} 
to a superstar-like tree (superhub) decorated by the hierarchy of trees (hubs), representing the market 
structure during the period of the crash. Subsequently, we found the transition from this complex tree to the 
power law tree decorated by the hierarchy of local star-like trees or hubs (where the richest from these hubs  
could be a candidate for another superhub) representing the market structure and topology after the worldwide 
financial crash. 

We foresee that our results, being complementary to others obtained earlier \cite{FFH,FFHol,DroKwap,KwDr}, 
can serve as a phenomenological foundation for the modeling of dynamic structural and topological phase transitions 
and critical phenomena on financial markets \cite{WH}.

\section{Results and discussion}

\begin{figure}
\begin{center}
\includegraphics[width=120mm,angle=0,clip]{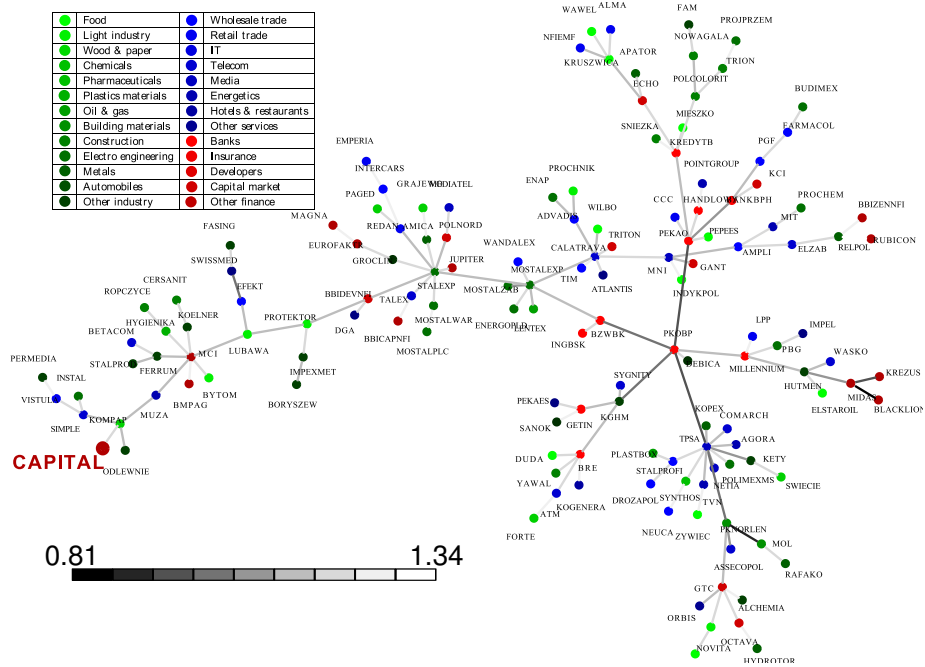} 
\caption{The hierarchical MST associated with the WSE (and consisting of $N=142$ companies) for the period from 
2005-01-03 to 2006-03-09, before the 
worldwide financial crash. The companies are indicated by the coloured circles (see the legend for an additional 
description). We focus on the financial company CAPITAL Partners (large red circle), as later it plays a central 
role in the MST, shown in Figure \ref{figure:20080812_asien}. When the link between two companies is in dark grey, 
the cross-correlation between them is greater, while the distance between them is shorter (cf. the corresponding 
scale incorporated there). However, the geometric distances between companies, shown in the Figure by the lengths 
of straight lines, are arbitrary, otherwise the tree would be much less readable.}
\label{figure:20060309_asien}
\end{center}
\end{figure}

The initial state (graph or complex network) of the WSE is shown in Figure \ref{figure:20060309_asien} in the form 
of a hierarchical MST\footnote{For the construction of the MST network, we here used Prim's algorithm \cite{PA}, which 
is quicker than Kruskal's algorithim \cite{PA,JK}, in particular for the number of companies $N\gg 1$. Both algorithms 
are often used.}. 
This graph was calculated for $N=142$ companies present on WSE for the period from 2005-01-03 to 
2006-03-09, i.e. before the worldwide financial crash occurred \cite{DiSo}. 

\begin{figure}
\begin{center}
\includegraphics[width=120mm,angle=0,clip]{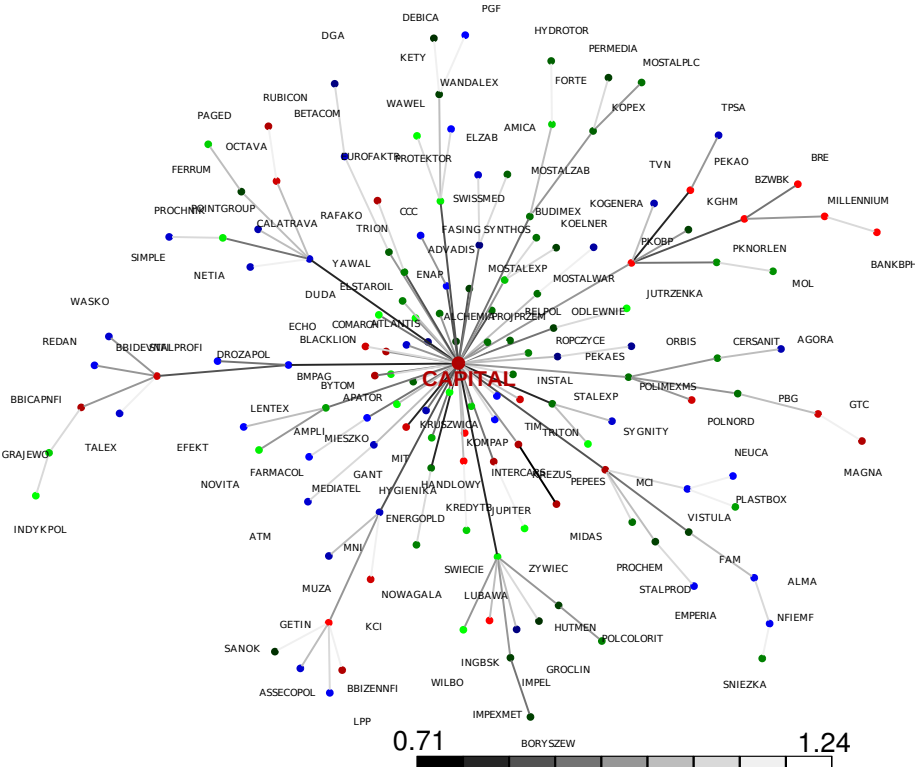} 
\caption{The superstar-like graph (or superhub) of the MST (also consisting of $N=142$ companies 
of the WSE) observed for the period from 2007-06-01 to 2008-08-12, which covers the worldwide financial crash. 
Now CAPITAL Partners becomes a dominant hub (or superhub). It is a temporal giant component, i.e. the central company 
of the WSE.}
\label{figure:20080812_asien}
\end{center}
\end{figure}

We focus on the financial company CAPITAL Partners\footnote{The full name of this company is CAPITAL Partners. It 
was listed on the WSE from 20 October 2004. The main activities of the company are capital investment in various 
assets and investment advice.}. It is a suburban company for the most of the period in question. However, it 
becomes a central company for the MST presented in Figure \ref{figure:20080812_asien}, for the period from 
2007-06-01 to 2008-08-12, which covers the worldwide financial crash.

In other words, for this period of time, CAPITAL Partners is represented by a vertex which has a much larger number
of edges (or it is of a much larger degree) than any other vertex (or company). This means that it becomes 
a dominant hub (superhub) or a giant component. 

In the way described above, the transition between two structurally (or topologically) different states of the 
stock exchange is realized. We observed the transition from hierarchical (power law) MST (consisting of a hierarchy
of local stars or hubs) to the superstar-like (or superhub) MST decorated by the hierarchy of trees (hubs).

\begin{figure}
\begin{center}
\includegraphics[width=120mm,angle=0,clip]{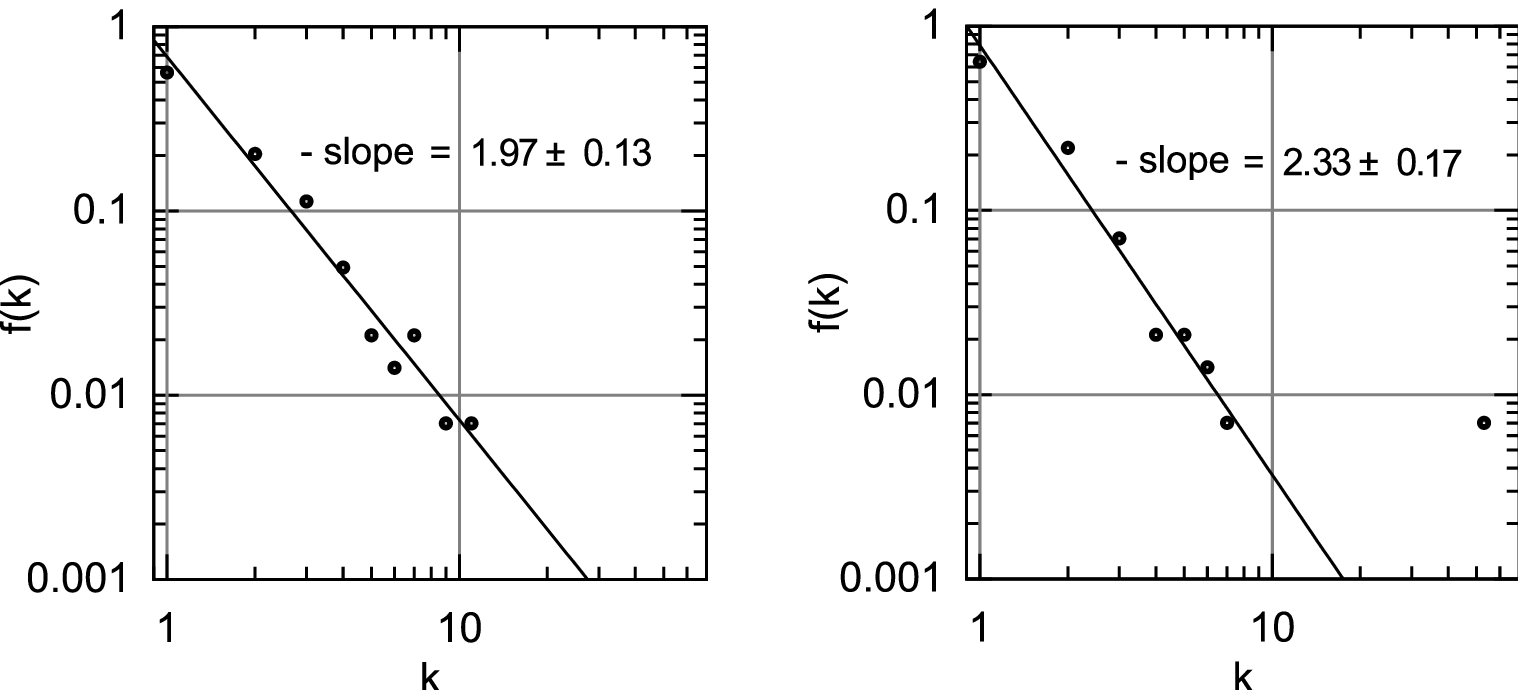} 
\caption{The comparison of power law discrete distributions $f(k)$ vs. $k$ (where $k$ is the vertex degree) for the 
hierarchical MST shown in Figure \ref{figure:20060309_asien} and the superstar-like MST decorated by the hierarchy 
of trees shown in Figure \ref{figure:20080812_asien}. One can observe that for the latter MST there is a single 
vertex (rhs plot), which has a degree much larger (equalling 53) than any other vertex. Indeed, this vertex 
represents the company CAPITAL Partners, which seems to be a superextreme event or a dragon king 
\cite{DSDS,WGKS,AJK,MaSo}, being a giant component of the MST network \cite{DGM}.}
\label{figure:spok_nie_spok}
\end{center}
\end{figure}

In Figure \ref{figure:spok_nie_spok} we compare discrete distributions of vertex degrees\footnote{The 
discrete distribution of the vertex degree is normalized by a factor equal to the total number of vertices $N$ fixed 
for a given period of time.}.

Although the distributions obtained are power laws, we cannot say that we are here dealing with 
a Barab{\'a}si--Albert (BA) type of complex network with their rule of preferential linking of new vertices 
\cite{AB}. This is because for both our trees, the power law exponents are distinctly smaller than 3 (indeed, the 
exponent equal to 3 characterizes the BA network), which is a typical observation for many real complex networks 
\cite{DGM}. 

%The slope (equalling $-2.33\mp 0.17$) of the rhs plot in this figure.
% is also characteristic of the complex network of actors (where superstars are also present) \cite{WS,ASBS}.
Remarkably, the rhs plot in Figure \ref{figure:spok_nie_spok} makes it, perhaps, possible to consider the tree presented in 
Figure \ref{figure:20080812_asien} as a power law MST decorated by a temporal dragon king\footnote{The equivalent 
terms `superextreme event' and `dragon king' stress that \cite{DSDS}: (i) we are dealing with an exceptional event 
which is completely different in comparison with the usual events; (ii) this event is significant, being distinctly 
outside the power law. For instance, in paper \cite{WGKS} the sustained and impetuous dragon kings were defined and 
discussed.}. This is because the single vertex (representing CAPITAL Partners) is located far from the straight
line (in the log-log plot) and can be considered as a temporally outstanding, superextreme event or a temporal dragon 
king \cite{DSDS,WGKS,AJK,MaSo}, which condenses the most of the edges (or links). Hence, the probability 
$f(k_{max})=0.007=1/142$, where $k_{max}=53$ is the degree of the dragon king (which is the maximal degree here). 
We suggest that the appearance of such a dragon king could be a signature of a crash\footnote{Obviously, this is 
a promising hypothesis which requires, however, a more systematic study.}. 

\begin{figure}
\begin{center}
\includegraphics[width=120mm,angle=0,clip]{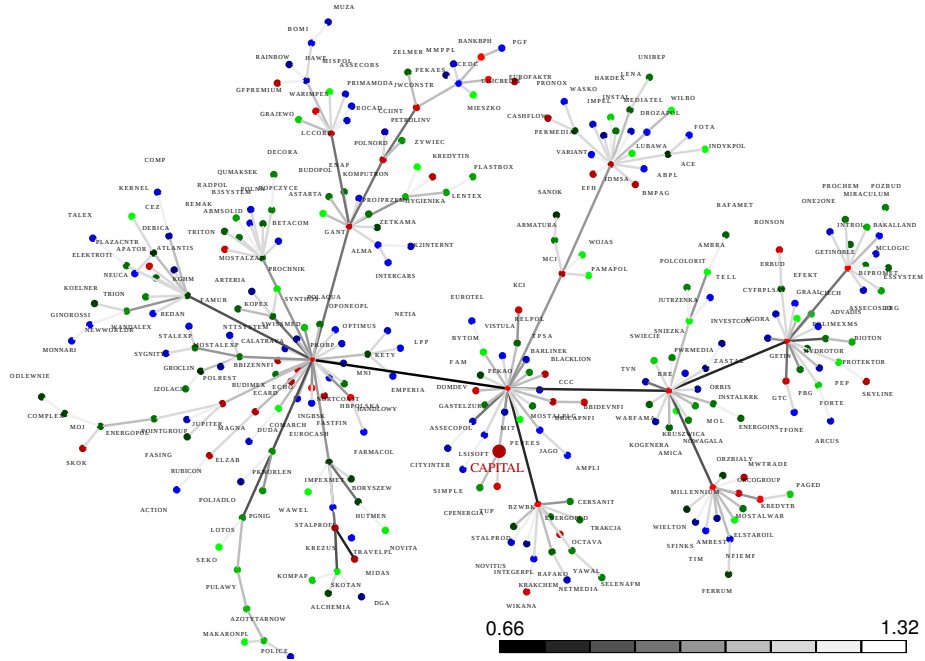}
\caption{The hierarchical graph of the MST decorated by several local star-like trees for the WSE for the period 
from 2008-07-01 to 2011-02-28, that is after the worldwide financial crisis. The companies are indicated by 
colored circles (see the legend in Figure \ref{figure:20060309_asien}).
Apparently, CAPITAL company is no longer the central hub, but has again become a marginal company (vertex). 
When the link between two companies is in dark gray, the cross-correlation between them is greater, while the 
distance between them is shorter. However, the geometric distances between companies, shown in the figure by the 
length of the straight lines, are arbitrary, otherwise the tree would be much less readable.}
\label{figure:20080701-20110228_New}
\end{center}
\end{figure}

\begin{figure}
\begin{center}
\includegraphics[width=120mm,angle=0,clip]{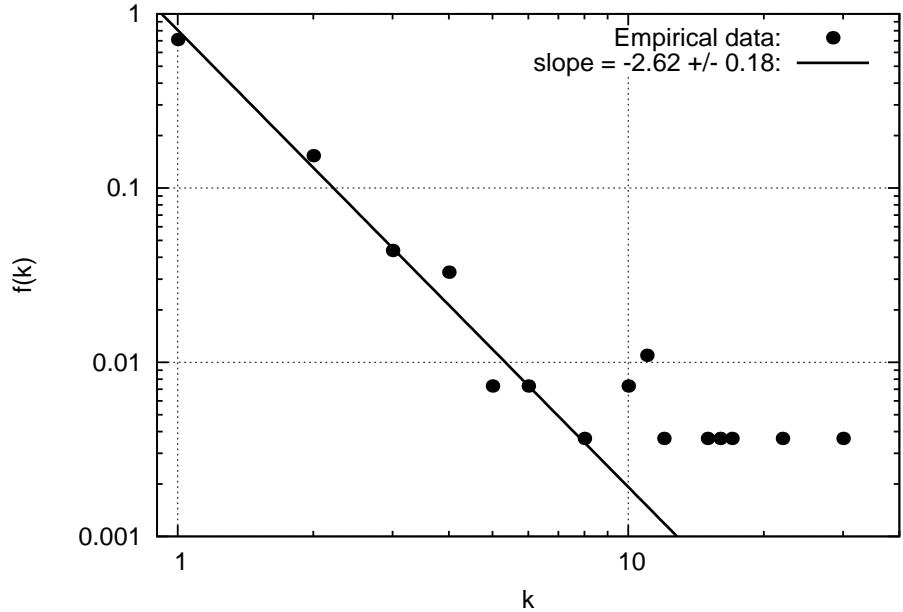} 
\caption{The power law discrete distribution $f(k)$ vs. $k$ (where $k$ is the vertex degree) for the MST shown in 
Figure \ref{figure:20080701-20110228_New}. Six points (associated with several different companies) appeared above 
the power law. This means that several large hubs appeared instead of a single superhub. Apparently, the richest 
vertex has here the degree $k_{max}=30$ and the corresponding probability $f(k_{max})=0.0036=1/274$. However, this 
vertex cannot be considered as a superextreme event (or dragon king) because it is not separated far enough from 
other vertices.}
\label{figure:214_New}
\end{center}
\end{figure}

For completeness, the MST was constructed for $N=274$ companies of the WSE for a third period of time, 
from 2008-07-01 to 2011-02-28, i.e. after the worldwide financial crash (cf. Figure 
\ref{figure:20080701-20110228_New}). 
It is interesting that several new (even quite rich) hubs appeared while the single superhub (superstar) disappeared 
(as it became 
a marginal vertex). This means that the structure and topology of the network strongly varies during its evolution
over the market crash. This is also well confirmed by the plot in Figure \ref{figure:214_New}, where several 
points (representing large hubs) are located above the power law. Apparently, this power law is 
defined by the slope equal to $-2.62 \mp 0.18$ and cannot be considered as a BA complex network. Rather, it is 
analogous to the internet, which is characterized by almost the same slope \cite{FFF,CCGJSW}. It would be an 
interesting project to identify the actual local dynamics (perhaps nonlinear) of our network, which subsequently 
creates and then annihilates the temporal singularity (i.e. the temporal dragon king).

\begin{figure}
\begin{center}
\includegraphics[width=120mm,angle=0,clip]{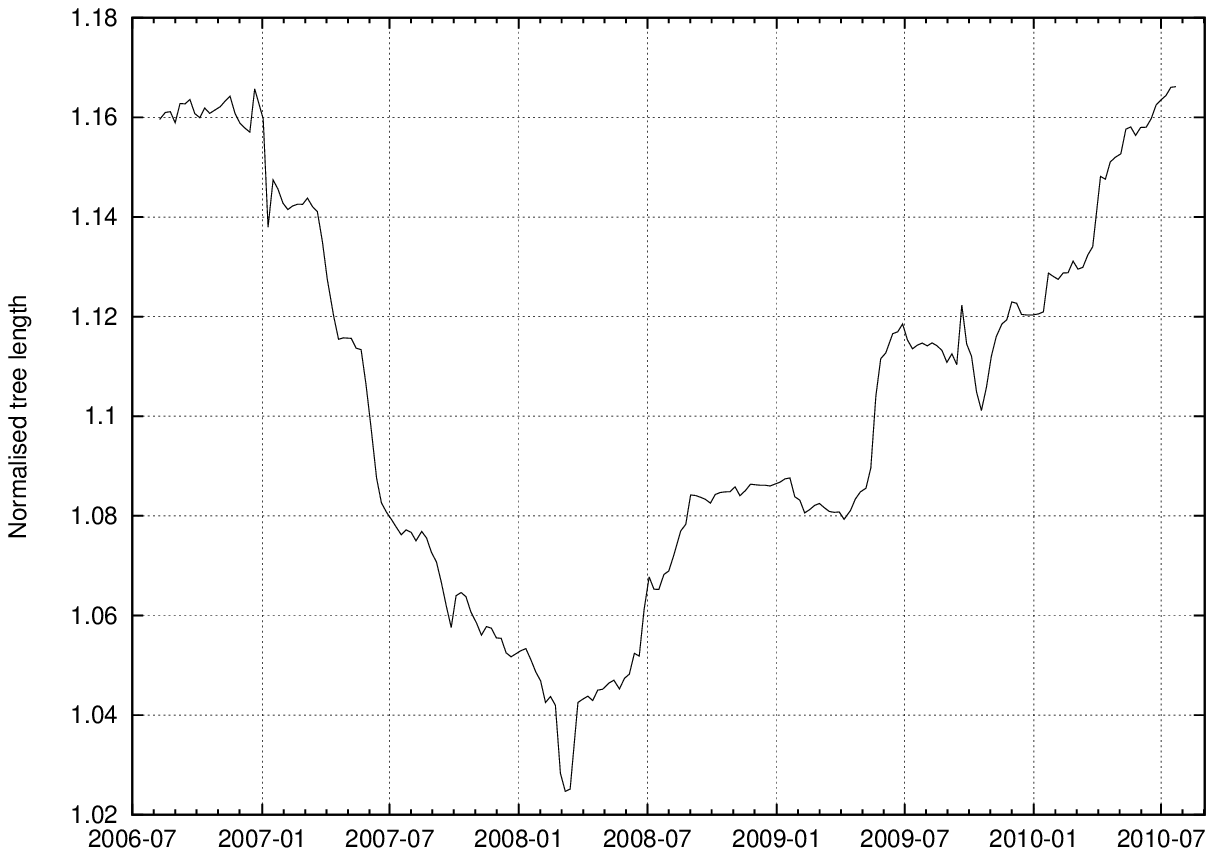} 
\caption{Normalized length of the MST vs. time (counted in trading days (td)). Apparently, the well-defined 
absolute minimum of the curve is located at the beginning of 2008. This localization 
(in the period from 2007-06-01 to 2008-08-12, covering the crash), together with the corresponding length so close 
to 1.0, confirm the existence of a network, which is significantly more compact than others.}
\label{figure:tree_length}
\end{center}
\end{figure}

\begin{figure}
\begin{center}
\includegraphics[width=120mm,angle=0,clip]{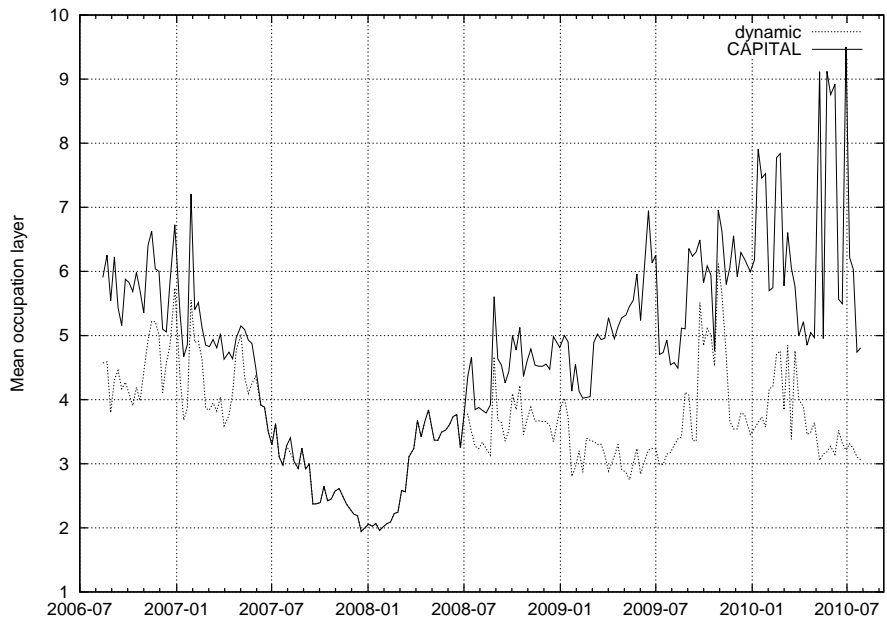} 
\caption{Mean occupation layer for the MST vs. time (counted in trading days, (td)), where CAPITAL Partners 
was assumed to be the central hub (the solid curve). For comparison, the result based on the central temporal hubs 
(having currently the largest degrees, the dotted curve) was obtained. 
%This dashed curve is termed dynamic.
Apparently, the well-defined absolute minimum, common for both curves, is located at the beginning of 2008 (in the 
period from 2007-06-01 to 2008-08-12).}
\label{figure:mean_layer}
\end{center}
\end{figure}

The considerations given above are confirmed in the plots shown in Figures \ref{figure:tree_length} and 
\ref{figure:mean_layer}. There well-defined absolute minima of the normalized length and mean occupation layer 
vs. time at the beginning of 2008 are clearly shown, respectively.  

As usual \cite{OCKK1,OCKK2,OCKKK}, the normalized length of the MST network simply means the average length of the 
edge directly connecting two vertices\footnote{Obviously, the edge between two vertices is taken into account only if 
any connection between them exists.}. 
%The normalization was chosen so as to obtain length equals 1 for a pure star-like tree. 
Apparently, this normalized length vs. time has an absolute minimum close to 1 at the beginning
of 2008 (cf. Figure \ref{figure:tree_length}), while at other times much shallower (local) minimums are observed. 
This result indicates the existence of a more compact structure at the beginning of 2008 than at 
other times. 

Furthermore, by applying the mean occupation layer defined, as usual \cite{OCKK1,OCKK2,OCKKK}, by the mean number 
of subsequent edges connecting a given vertex of a tree with the central vertex (here CAPITAL Partners), we obtained 
quite similar results (cf. the solid curve in Figure \ref{figure:mean_layer}). For comparison, the result based 
on the other central temporal hubs (having currently the largest degrees) was also obtained (cf. the dotted curve
in Figure \ref{figure:mean_layer}). This approach is called the dynamic one. Fortunately, all the approaches used above 
give fully consistent results which, however, require some explanation. 

In particular, both curves in 
Fig. \ref{figure:mean_layer} coincide in the period from 2007-06-01 to 2008-08-12 having common abolute minimum 
located at the beginning of 2008. To plot the dotted curve the company, which has the largest degrees, was chosen at 
each time as a temporal central hub. In general, such a company can be replaced from time to time by other company. 
However, for the period given above indeed the CAPITAL Partners has largest degrees (while other companies have 
smaller ones, of course). This significant observation is clearly confirmed by the behavior of the solid curve 
constructed at a fixed company assumed as a central hub, which herein it is the CAPITAL Partners. Just outside this 
period, the CAPITAL Partnes is no more a central hub (becoming again the peripheral one) as other companies play then 
his role, although not so spectacular. This results from the observation that the dotted curve in 
Fig. \ref{figure:mean_layer} is placed below the solid one outside the second period (i.e. from 2007-06-01 to 
2008-08-12). Hence, we were forced 
to restrict the period on August 12, 2008 and do not consider other period such as between September 2008 and March 
2009, where the most serious drawdown during the worldwide financial crash 2007-2009 occurred\footnote{However, our 
complementary calculations for the Frankurt Stock Exchange support our approach. The role of central hub (or superhub) 
plays there the GITTARSALZ AG Stahl und Technologie company.}. Perhaps, some precursor of the crash is demonstrated 
herein by the unstable state of the WSE. Anyway, the subsequent work should contain a more detailed analysis of the 
third period.

The existence of the absolute minimum (shown in Figure \ref{figure:mean_layer}) for CAPITAL Partners, and 
simultaneously the existence of the absolute minimum shown in Figures \ref{figure:tree_length} in the first 
quarter of 2008 (to a satisfactory approximation) confirms the existence of the star-like structure (or a superhub), 
as a giant component of the MST, centered around CAPITAL Partners. We may suppose that the evolution from 
a marginal to the central company of the stock exchange and again to a marginal company, is stimulated perhaps 
by the most attractive financial products offered by this company to the market only in the second period of time 
(i.e. in the period from 2007-06-01 to 2008-08-12).

\section{Concluding remarks}

In this work, we have studied the empirical evolving connected correlated network associated with 
a small size stock exchange, the WSE. Our result seem somewhat embarrassing that such a marginal capitalization 
company as CAPITAL (less than one permil of a typical WIG20 company, like for instance KGHM) becomes a dominant 
hub (superhaub) in the second period considered (see Fig. 2 for details). 

%{\color{blue} One of the possible explanation is based 
%on the hypothesis that crisis makes relations between companies of real economy weaker (or distance between them makes 
%longer) than their relations with institutions from financial sector. That is, even weak relations with financial 
%institutions before the crisis can become relatively significant during the duration time of the crisis. Hence, 
%the Minimal Spanning Tree algorithm chooses, instead of direct distances between these companies, the indirect 
%one that is, over the financial institution (e.g. such as the CAPITAL Partners one) since the distance to this 
%institution can be respectively shorter now. In this way the super-star like structure can be constructed, even around 
%the financial institution of quite low capitalization. A typical example could be the STALEXP and MOSTALEXP companies 
%(placed in the middle part of MST shown in Fig. 1). Apparently, they were directly linked within the first period, 
%i.e. before the crisis. However, the crisis changed this structure making link between them indirect that is, over 
%the CAPITAL Partners (see Fig. \ref{figure:20080812_asien} for details).} 
 
Our work provides an empirical evidence that there is a dynamic structural and topological the first 
order phase transition in the time range dominated by a crash. Namely, before and after this range the superhub 
(or the unstable state of the WSE) 
disappears and we observe the power law MST and power law MST decorated by several hubs\footnote{The best 
candidate for the superhub within the third period (from 2008-07-01 to 2011-02-28) could be, perhaps, PKOBP (see 
Figs. \ref{figure:20080701-20110228_New} and \ref{figure:214_New} for details), which is a richest vertex having degree 
equals 30.}, respectively. Therefore, our results consistently confirm the existence of the dynamic structural and 
topological phase transitions, which can be roughly summarized as follows:
\begin{eqnarray}
& &\mbox{phase of power law MST - a stable state} \nonumber \\ 
& &\mbox{$\Rightarrow $ phase of the superstar-like or superhub MST decorated by hierarchy} \nonumber \\
& &\; \; \; \; \; \mbox{of trees or hubs - perhaps an unstable state} \nonumber \\
& &\mbox{$\Rightarrow $ phase of power law MST decorated by several star-like trees or hubs} \nonumber \\
& &\; \; \; \; \; \mbox{(where the richest hub could be a candidate for another superhub)} \nonumber \\
& &\; \; \; \; \; \mbox{- perhaps a stable state}. \nonumber 
\end{eqnarray}
We assume the hypothesis that the first transition can be considered as a signature of a stock exchange crash, while 
the second one can be understood to be an aftershock. Nevertheless, the second transition related to 
the third period requires a more detailed analysis. Indeed, in this period the PKOBP to much resemble a superhub
(see Figs. \ref{figure:20080701-20110228_New} and \ref{figure:214_New} for details), which could play a role of other 
stable state of the WSE. In other words, our work indicates that we deal perhaps with indirect transition 
(the first order one) between two stable components, where the unstable component is surprisingly well seen among them. 

One of the most significant observations contained in this work comes from plots in Figures 
\ref{figure:spok_nie_spok} and \ref{figure:214_New}. Namely, the exponents of all degree distributions are smaller 
than 3, which means that all variances of vertex degrees diverge.
%\footnote{It seems that a degree distribution exponent even smaller than 2 could occur for the MST obtained for the 
%period from 2005-01-03 to 2006-03-09 (cf. the plot in Figure \ref{figure:20060309_asien}).}. 
This indicates that we are here dealing with criticality 
as the range of fluctuations is compared with the size of the graph. This means that the network evolution from 
2005-01-03 to 2011-020-028 takes place within the scaling region\footnote{Note that the scaling region it is a region 
where both the first order and signatures of the second order phase transitions are present together. We suppose herein 
that this is, indeed, our case.} \cite{DGM,DS0,BaPo,HH} containing a critical point. 
Apparently, we are here dealing with scale-free networks, which are ultrasmall worlds \footnote{For the 
ultrasmall world the mean length between two vertices of a graph is proportional to $\ln \ln N$ instead of that for 
the small world proportional only to $\ln N$.} \cite{RCSH,WS,ASBS}. It should be stressed that similar 
results we also obtained for Frankfurt Stock Exchange\footnote{In fact, we obtained results analogous to those 
presented in all our Figs. 1--7.}. 

We suppose that our results are complementary to those obtained earlier by Dro\.zd\.z and Kwapie\'n
\cite{DroKwap}. Their results focused on the slow (stable) component (state). Namely, they constructed the MST network 
of 1000 highly capitalized American companies. The topology of this MST show its centralization around the most 
important quite stable node being the General Electric. This was found both in the frame of binary and weighted MSTs. 

Noteworthly, the fact should be stressed in this context that the discontinuous phase transition (i.e. 
the first order phase one) evolves continuously before the continuous phase transition (i.e. before the second order 
one). This discontinuous phase transition goes over the unstable state involving, perhaps, a superheating  
state such as the superhub in our case. This cannot be considered as a noise\footnote{Indeed, the case of 
the noise was discussed in this context in paper \cite{STZM}.} in the system but rather should be considered as 
a result of the natural evolution of the system until the critical point is reached (cf. \cite{KwDr} and refs. therein, 
where the role of stable states (or slow components) on NYSE or NASDAQ was considered by using binary and weighted 
MSTs).

We suppose that the phenomenological theory of cooperative phenomena in networks proposed by Goltsev 
et al. \cite{GDM} (based on the concepts of the Landau theory of continuous phase transitions) could be a promising 
first attempt.

An alternative view of our results could consider the superhub phase as a temporal condensate \cite{DGM} (and refs. 
therein). Hence, we can reformulate the phase transitions mentioned above as representing the dynamic transition 
from the disordered phase into a temporal condensate, and then the transition from the condensate again to some 
disordered phase.

We hope that our work is a good starting point to find similar 
topological transitions at other markets. For instance, we also studied a medium size stock exchange, the Frankfurt 
Stock Exchange. Because the results obtained resemble very much those found for the Warsaw Stock Exchange, we omitted 
them here. Furthermore, the analytical treatment of the dynamics of such a network remains a challenge.

We can summarize this work with the conclusion that it could be promising to study in details the phase transitions 
considered above, which can define the empirical basis for understanding of stock market evolution as a whole. 

\section*{Acknowledgments}

We are grateful Rosario N. Mantegna and Tiziana Di Matteo for helpful comments and suggestions.

\end{document}